\documentclass[%superscriptaddress,
preprint,
showpacs,preprintnumbers,
bibnotes,
 amsmath,amssymb,
 aps,
 pra,
superscriptaddress,
longbibliography,
]{revtex4-1}

\usepackage{amsmath}
\usepackage{amsfonts}
\usepackage{amssymb}
\usepackage{graphicx}
\usepackage{xspace}
\usepackage{color}
\usepackage{comment}
\usepackage[hidelinks]{hyperref}
\usepackage{dcolumn}% Align table columns on decimal point
\usepackage{bm}% bold math
\usepackage{mathtools}

\usepackage{comment}

\usepackage[detect-weight=true, detect-family=true]{siunitx}

\graphicspath{ {./figures/} }

\begin{document}

\title{
Tip-induced strain, bandgap, and radiative decay engineering of a single metal halide perovskite quantum dot
}
\author{Hyeongwoo Lee$^{\dagger}$}
\affiliation
{Department of Physics, Ulsan National Institute of Science and Technology (UNIST), \\Ulsan 44919, Republic of Korea}
\author{Ju Young Woo$^{\dagger}$}
\affiliation
{Manufacturing Process Platform R$\&$D Department, Korea Institute of Industrial Technology (KITECH), \\Ansan 15588, Republic of Korea}
\author{Dae Young Park}
\affiliation
{Department of Energy Science, Sungkyunkwan University (SKKU), \\Suwon 16419, Republic of Korea}
\author{Inho Jo}
\affiliation
{Department of Physics, Korea Advanced Institute of Science and Technology (KAIST), \\Daejeon 34141, Republic of Korea}
\author{Jusun Park}
\affiliation
{Department of Energy Science, Sungkyunkwan University (SKKU), \\Suwon 16419, Republic of Korea}
\author{Yeunhee Lee}
\affiliation
{Department of Physics, Korea Advanced Institute of Science and Technology (KAIST), \\Daejeon 34141, Republic of Korea}
\author{Yeonjeong Koo}
\affiliation
{Department of Physics, Ulsan National Institute of Science and Technology (UNIST), \\Ulsan 44919, Republic of Korea}
\author{Jinseong Choi}
\affiliation
{Department of Physics, Ulsan National Institute of Science and Technology (UNIST), \\Ulsan 44919, Republic of Korea}
\author{Hyojung Kim}
\affiliation
{Department of Energy Science, Sungkyunkwan University (SKKU), \\Suwon 16419, Republic of Korea}
\author{Yong-Hyun Kim}
\affiliation
{Department of Physics, Korea Advanced Institute of Science and Technology (KAIST), \\Daejeon 34141, Republic of Korea}
\author{Mun Seok Jeong}
\affiliation
{Department of Energy Science, Sungkyunkwan University (SKKU), \\Suwon 16419, Republic of Korea}
\author{Sohee Jeong$^{\ast}$}
\affiliation
{Department of Energy Science, Sungkyunkwan University (SKKU), \\Suwon 16419, Republic of Korea}
\author{Kyoung-Duck Park$^{\ast}$}
\affiliation
{Department of Physics, Ulsan National Institute of Science and Technology (UNIST), \\Ulsan 44919, Republic of Korea}
\email{kdpark@unist.ac.kr}
\date{\today}

\begin{abstract}

\noindent 
\textbf{
Strain engineering of perovskite quantum dots (pQDs) enables widely-tunable photonic device applications. 
However, manipulation at the single-emitter level has never been attempted.
Here, we present a tip-induced control approach combined with tip-enhanced photoluminescence (TEPL) spectroscopy to engineer strain, bandgap, and emission quantum yield of a single pQD.
Single CsPbBr$_{x}$I$_{3-x}$ pQDs are clearly resolved through hyperspectral TEPL imaging with $\sim$10 nm spatial resolution. 
The plasmonic tip then directly applies pressure to a single pQD to facilitate a bandgap shift up to $\sim$62 meV with Purcell-enhanced PL quantum yield as high as $\sim$10$^5$ for the strain-induced pQD.
Furthermore, by systematically modulating the tip-induced compressive strain of a single pQD, we achieve dynamical bandgap engineering in a reversible manner.
In addition, we facilitate the quantum dot coupling for a pQD ensemble with $\sim$0.8 GPa tip pressure at the nanoscale.
Our approach presents a new strategy to tune the nano-opto-electro-mechanical properties of pQDs at the single-crystal level.
}

\end{abstract}

\maketitle

\noindent 
Metal halide perovskite quantum dots (pQDs) are promising candidates for various optoelectronic device applications.
Specifically, their excellent optical and electrical properties, such as high photoluminescence (PL) quantum yields (PLQY), widely tunable bandgap, narrow emission width, high absorption coefficient, and high defect tolerance, enable a quantum leap in light-emitting \cite{liu2020, stranks2015, dong2020, swarnkar2015, dirin2016} and photovoltaic devices \cite{swarnkar2016, han2019}. 
To improve the performance of perovskite-based optoelectronic devices, many strategies have been proposed, such as composition engineering, interface modifications, and device structure engineering. 
As another of these approaches, strain engineering is an attractive method for improving device performance significantly by reversibly and widely tuning the optoelectronic properties of perovskites. 

Although a range of strain engineering techniques, including hydrostatic pressurization \cite{liu2018, ma2018}, electrostriction \cite{chen2018}, annealing \cite{steele2019, xiao2014}, and mechanical bending \cite{zhu2019, chen2020}, have been studied, controlling strain at the nanoscale or single-crystal level has never been attempted due to the difficulties of local strain control as well as \textit{in-situ} nano-optical characterizations. 
In addition, to widely tune the bandgap or even surpass the threshold of the bond-breaking energy in nanocrystals, a high pressure at the $\sim$GPa scale is generally required \cite{liu2019, liu2018}. 
However, applying hydrostatic pressure of $\sim$GPa scale and maintaining it in device platforms is highly challenging, which restricts its practical applications. 
Hence, to understand the fundamental nano-optomechanical properties of single nanocrystals, control their properties at the nanoscale, and further adopt these advantages in practical applications, strain engineering at the single-nanocrystal level is highly desirable.

In this work, we present a tip force control approach combined with tip-enhanced photoluminescence (TEPL) spectroscopy to engineer the strain of single CsPbBr$_{x}$I$_{3-x}$ pQDs while simultaneously investigating their modifying emission behaviors \cite{park2019}. 
Using Purcell enhanced TEPL signal by a factor of $\sim$10$^{5}$, we probe single CsPbBr$_{x}$I$_{3-x}$ pQDs with a spatial resolution of $\sim$10 nm. 
We then systematically press the single pQDs through atomic force microscopy (AFM) control of an Au tip. 
The applied compressive strain to a single pQD gives rise a gradual bandgap redshift up to $\sim$62 meV as well as dramatic enhancement of PLQY by the Purcell effect in the plasmonic nano-cavity, which are quantitatively explained by theoretical density functional theory (DFT) calculations and finite-difference time domain (FDTD) simulations \cite{lee2020, purcell1946, schuller2010, russell2012}. 
Furthermore, the bandgap energy of a single pQD is gradually shifted back to the original state when we retract the Au tip from it, i.e., systematic bandgap engineering of a single pQD is achieved in a reversible manner. 
This dynamic control is achieved by exploiting the tip indentation region of $\sim$10 nm$^2$, which enables local pressure up to $\sim$0.8 GPa. 
In addition, with this extremely high local pressure, we facilitate the quantum dot coupling in a nanoscale region of the CsPbBr$_{x}$I$_{3-x}$ pQD ensemble characterized by structural distortion and spectral redshift up to $\sim$8 meV \cite{williams2009, cui2019, li2019}. 
\\

%%%%%%%%%%%%%%%%%%%%%%%%%%%%%%%%%%%%%%%%%%%%%
\begin{figure*}
	\includegraphics[width = 16 cm]{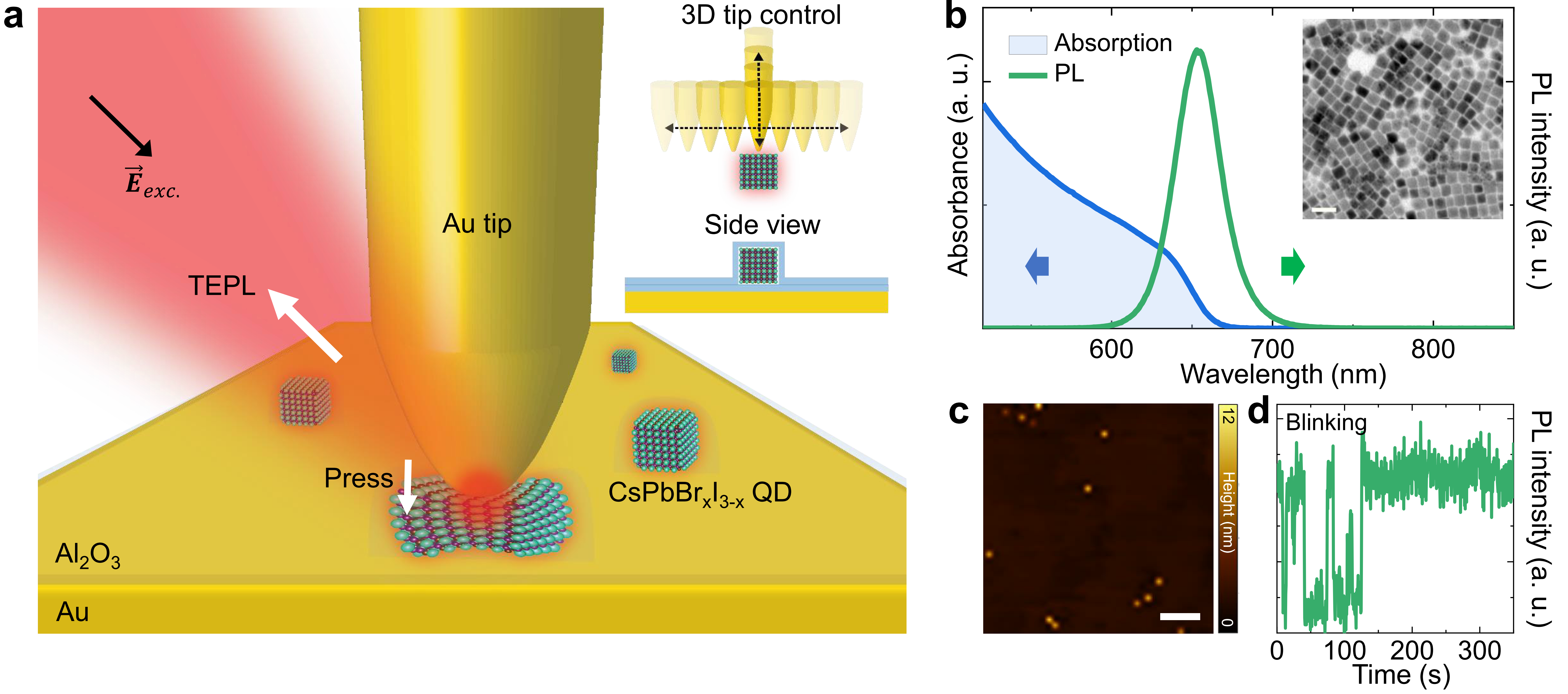}
	\caption{
\textbf{Schematic of experimental setup and characterization of single pQDs.} 
(a) Schematic illustration of TEPL spectroscopy to probe and control the radiative emission of single CsPbBr$_{x}$I$_{3-x}$ pQDs. 
Individually distributed CsPbBr$_{x}$I$_{3-x}$ pQDs are encapsulated by an Al$_{2}$O$_{3}$ layer (0.5 nm thickness). 
Three-dimensional tip-positioning and atomic force tip control allow single CsPbBr$_{x}$I$_{3-x}$ pQDs to be found and induce local pressure on them. 
(b) Absorption (blue) and PL spectra (green) of CsPbBr$_{x}$I$_{3-x}$ pQDs measured from the ensemble state. 
Inset: TEM image of drop-cast CsPbBr$_{x}$I$_{3-x}$ pQDs.
Scale bar is 40 nm. 
(c) AFM topography image of spin-coated CsPbBr$_{x}$I$_{3-x}$ pQDs on the Au substrate. 
Scale bar is 50 nm.
(d) Time-dependent far-field PL intensity of a few CsPbBr$_{x}$I$_{3-x}$ pQDs exhibiting a typical blinking behavior.   
}
	\label{fig:fig1}
\end{figure*}
%%%%%%%%%%%%%%%%%%%%%%%%%%%%%%%%%%%%%%%%%%%%% 

\noindent
\textbf{Pre-characterizations of single isolated CsPbBr$_{x}$I$_{3-x}$ pQDs}

\noindent
CsPbBr$_{x}$I$_{3-x}$ pQDs are spin-coated onto a template-stripped Au substrate with a thin dielectric layer (0.5 nm thick Al$_{2}$O$_{3}$). 
The single isolated pQDs are then covered again with an Al$_{2}$O$_{3}$ capping layer to prevent degradation under ambient conditions \cite{liu2019}. 
We use TEPL spectroscopy with a side-illumination geometry for nano-optomechanical characterizations of the single pQDs, as illustrated in Fig.~\ref{fig:fig1}a.
For the nanoscale positioning of the tip onto the single pQDs and vertical distance control between the tip and sample with a precision of $<$0.2 nm, the TEPL setup is designed based on shear-force AFM \cite{park2018}. 
We use an electrochemically etched Au tip (apex size of $\sim$15 nm) to obtain strong tip-enhanced PL responses of CsPbBr$_{x}$I$_{3-x}$ pQDs at an excitation wavelength of 632.8 nm. 
In addition, the Au tip can dynamically induce pressure on single CsPbBr$_{x}$I$_{3-x}$ pQDs through atomic force control at nanoscale (see Methods for more details). 
Before we prepare the single isolated CsPbBr$_{x}$I$_{3-x}$ pQDs, we conduct pre-characterizations of absorption, PL, and transmission electron microscopy (TEM) for the ensemble state to understand the fundamental optical and structural properties, as shown in Fig.~\ref{fig:fig1}b. 
The broad spectral range of absorption covers the excitation wavelength (632.8 nm) of our TEPL spectroscopy experiment.
The PL spectrum exhibits a sharp emission peak at $\sim$650 nm, and the TEM image verifies a uniform size distribution of CsPbBr$_{x}$I$_{3-x}$ pQDs.
We then dilute the solution and spin-coat it on the Au substrate to prepare the single isolated pQDs (see Methods for more details).
To characterize the distribution of pQDs, AFM measurement is performed, as shown in Fig.~\ref{fig:fig1}c.
The single CsPbBr$_{x}$I$_{3-x}$ pQDs are distributed well on the Au substrate for dynamic strain engineering of them individually.
Fig.~\ref{fig:fig1}d shows the change of far-field PL intensity of the CsPbBr$_{x}$I$_{3-x}$ pQDs over 350 s, exhibiting distinct blinking behaviors, which also indicates the small number of pQDs in the focused beam spot. 
\\

%%%%%%%%%%%%%%%%%%%%%%%%%%%%%%%%%%%%%%%%%%%%%
\begin{figure*}
	\includegraphics[width = 12 cm]{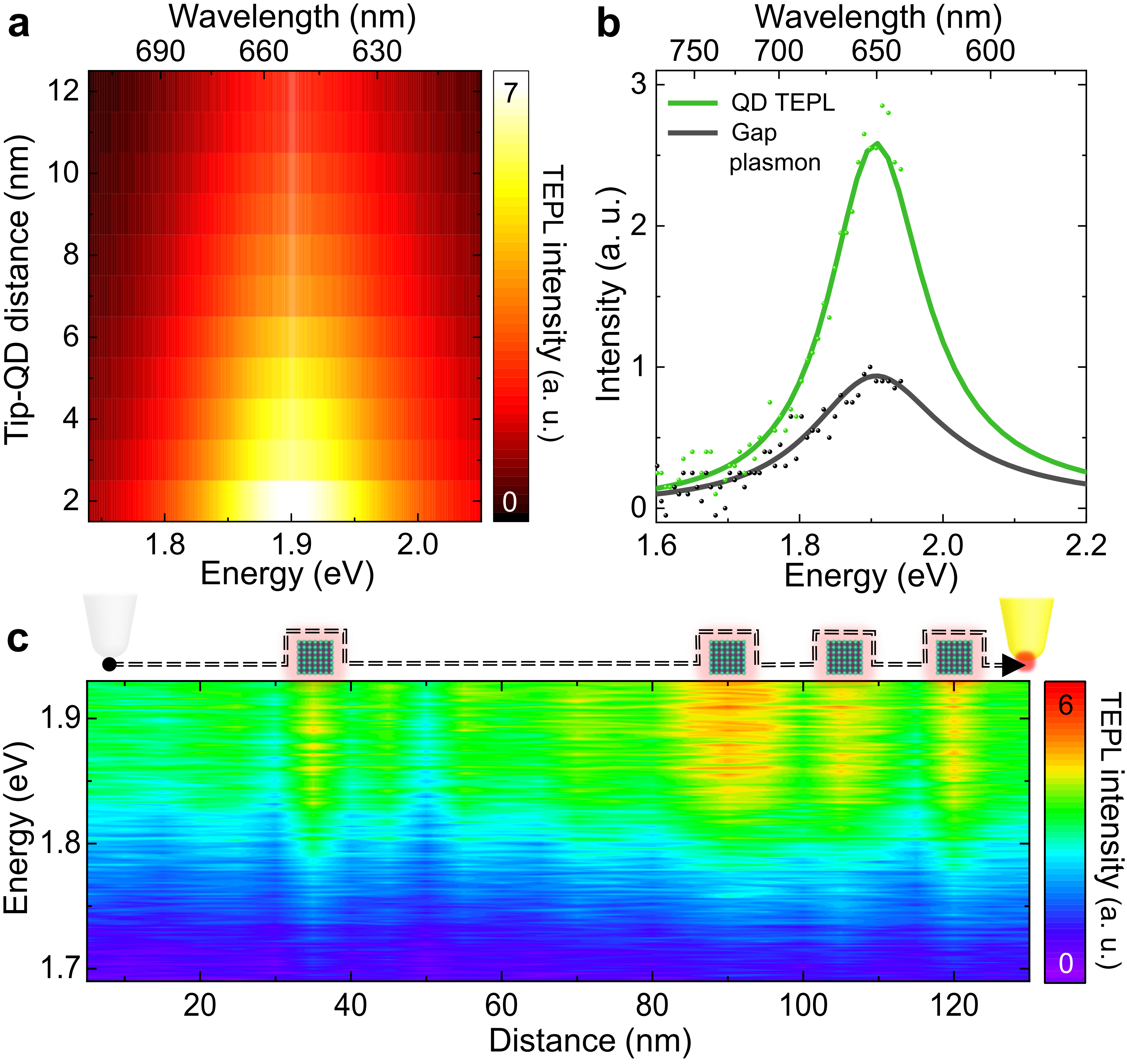}
	\caption{
\textbf{TEPL measurements of single pQDs.} 
(a) Evolution of TEPL spectra of single CsPbBr$_{x}$I$_{3-x}$ pQDs as a function of the distance between the Au tip and sample. 
(b) TEPL spectra (green) of single CsPbBr$_{x}$I$_{3-x}$ pQDs and the gap plasmon response (black) from the junction between the Au tip and Au substrate. 
The spectra in (a) and (b) are fitted with the Voigt function.
(c) Contour plot of TEPL spectra when the tip moves along the lateral direction of CsPbBr$_{x}$I$_{3-x}$ pQDs with a constant tip$\--$sample distance. 
}
	\label{fig:fig2}
\end{figure*}
%%%%%%%%%%%%%%%%%%%%%%%%%%%%%%%%%%%%%%%%%%%%%

\noindent
\textbf{TEPL measurements of the single isolated CsPbBr$_{x}$I$_{3-x}$ pQDs}

\noindent
To verify the tip-enhanced effect of the TEPL response as the plasmonic tip approaches the pQDs, we measure TEPL spectra as a function of distance, as shown in Fig.~\ref{fig:fig2}a. 
The evolution of the TEPL spectra shows a dramatically increasing emission intensity with decreasing distance. 
This result describes the near-field enhancement of the CsPbBr$_{x}$I$_{3-x}$ pQDs emission through plasmon$\--$exciton coupling in the nano-cavity \cite{gross2018, park2019}. 
Because we use CsPbBr$_{x}$I$_{3-x}$ pQDs as a model system to induce resonance coupling between the plasmon response and pQD exciton PL at $\sim$1.9 eV, we investigate their spectral difference by measuring the TEPL spectra with and without the CsPbBr$_{x}$I$_{3-x}$ pQDs, as shown in Fig.~\ref{fig:fig2}b.
Without the pQDs, only the gap-plasmon response is observed (black) through optical coupling between the Au tip and flat Au substrate. 
The broad linewidth of $\sim$235 meV is in good agreement with the results of previous studies \cite{vasily2014, park2019}. 
By contrast, as the Au tip is positioned on the single CsPbBr$_{x}$I$_{3-x}$ pQDs, the gap plasmon resonance disappears because a single CsPbBr$_{x}$I$_{3-x}$ pQD with a height of $\sim$10 nm is sandwiched between the Au tip and Au substrate.
Instead, the large TEPL signal of the single CsPbBr$_{x}$I$_{3-x}$ pQDs is observed (green) with a narrow linewidth of $\sim$170 meV. 

We then spatio-spectrally resolve the TEPL response of CsPbBr$_{x}$I$_{3-x}$ pQDs by laterally scanning the Au tip, as shown in  Fig.~\ref{fig:fig2}c.
In the lateral scanning of $\sim$130 nm, a few distinct spots exhibiting a strong TEPL response from CsPbBr$_{x}$I$_{3-x}$ pQDs are observed. 
They show different emission intensities which are possibly attributable to the different dipole orientations of pQDs (see Fig. S1 for more details).
It should be noted that the PL response of these single pQDs cannot be resolved by far-field PL measurement. 
\\

%%%%%%%%%%%%%%%%%%%%%%%%%%%%%%%%%%%%%%%%%%%%%
\begin{figure*}
	\includegraphics[width = 16 cm]{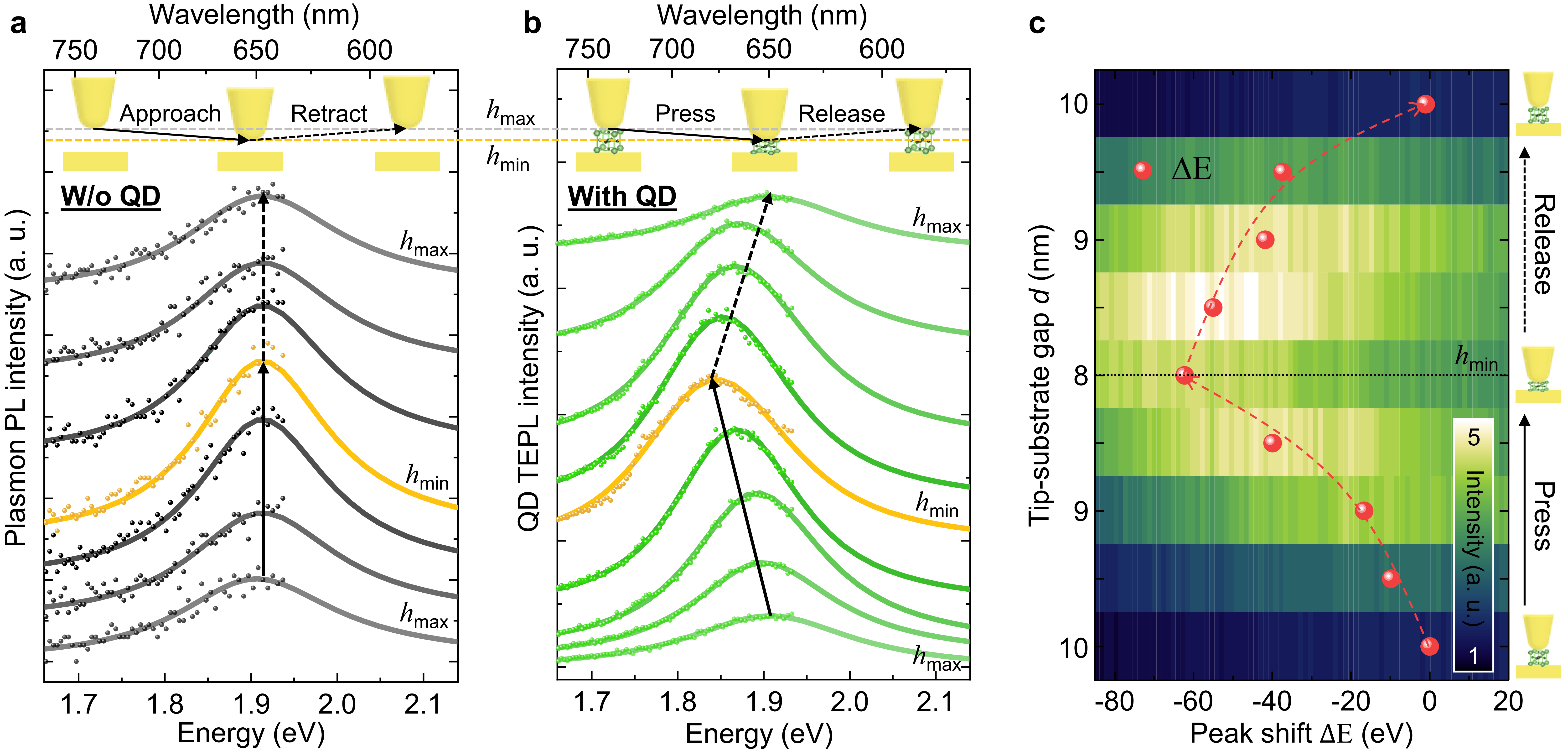}
	\caption{
\textbf{Strain, bandgap, and PLQY engineering of a single pQD.} 
(a) Spectra of the gap plasmon response when the Au tip approaches the Au substrate and retracts from it. 
(b) Reversible TEPL spectra of a single CsPbBr$_{x}$I$_{3-x}$ pQD with compressive stress changed by the tip. 
The spectra in (a) and (b) are fitted with the Voigt function. 
The orange curves in (a) and (b) show the spectra at the minimum height (h$_{min}$) and under maximum pressure, respectively.  
The relative height between the Au tip and Au substrate during the modulation process is the same as that for the gap plasmon, as in the illustrations at the top of (a) and (b).
(c) Peak shift (red) and intensity change (contour plot) of the TEPL spectra derived from (b). 
}
	\label{fig:fig3}
\end{figure*}
%%%%%%%%%%%%%%%%%%%%%%%%%%%%%%%%%%%%%%%%%%%%% 

\noindent
\textbf{Dynamic strain, bandgap, and quantum yield engineering of a single pQD}

\noindent
Beyond tip-enhanced spectroscopy, we then present tip-induced control of strain and optical properties of a single CsPbBr$_{x}$I$_{3-x}$ pQD. 
Before we perform the tip-induced strain engineering, we confirm the optical response of the gap plasmon when the distance between the Au tip and flat Au substrate is changed. 
We measure the evolution of the gap plasmon spectra through dynamic tip control in the vertical axis with respect to the Au surface. 
As shown in Fig.~\ref{fig:fig3}a, the plasmon intensity increases slightly as the gap decreases without a spectral shift.
Therefore, we can exclude the effect of gap plasmon in the bandgap engineering experiment for single pQDs. 
By constrast, distinct spectral redshifts are observed when the Au tip moves the same distance on a single CsPbBr$_{x}$I$_{3-x}$ pQD, as shown in Fig. 3b. 
From this AFM control, the Au tip can directly apply local pressure and induce compressive strain on a single pQD. 
When the tip is retracted, the pressure and strain applied on the single pQD are released.
This implies the capability of modulating the strain of a single pQD as well as the mode volume (V) of the plasmonic nano-cavity, which modifies the spontaneous emission rate of the pQD by the Purcell effect ($\propto$1$/$V). 
Hence, when we increases the tip pressure on the pQD, a gradual redshift of the TEPL peak caused by bandgap reduction is observed with increasing TEPL intensity. 
Interestingly, when we release the pressure by retracting the tip, the bandgap energy and emission intensity revert to their original state. 
As can be seen in Fig.~\ref{fig:fig3}c, robust bandgap and PLQY modulations are achieved in a reversible manner with a maximum energy shift of $\sim$62 meV.

In the situation of applying pressure larger than a certain value, a phase transition or an irreversible crystalline change has been reported \cite{liu2019, ma2018, liu2018}. 
To experimentally characterize this threshold pressure and prevent these irreversible changes to a CsPbBr$_{x}$I$_{3-x}$ pQD, we analyze the TEPL intensity change with respect to the tip pressure, as shown in Fig.~\ref{fig:fig3}c. 
When the perovskite crystals experience a gradual increase in pressure, the PL intensity should be decreased due to lattice distortion and bending of chemical bonds \cite{chen2020, liu2019}. 
In our results, because the field enhancement and Purcell effect are increased as the tip$\--$sample gap decreases, the PL emission of the CsPbBr$_{x}$I$_{3-x}$ pQD is enhanced until it reaches the threshold (\textit{d} $=$ 8.5 nm). 
However, from this point, because the degree of quenching exceeds the field enhancement, the intensity starts to decrease, as can be seen in Fig.~\ref{fig:fig3}c. 
This threshold is in good agreement with previous studies on the pressure-dependent intensity evolution of halide perovskite nanocrystals \cite{ma2018, liu2019}.
\\

%%%%%%%%%%%%%%%%%%%%%%%%%%%%%%%%%%%%%%%%%%%%%
\begin{figure*}
	\includegraphics[width = 16 cm]{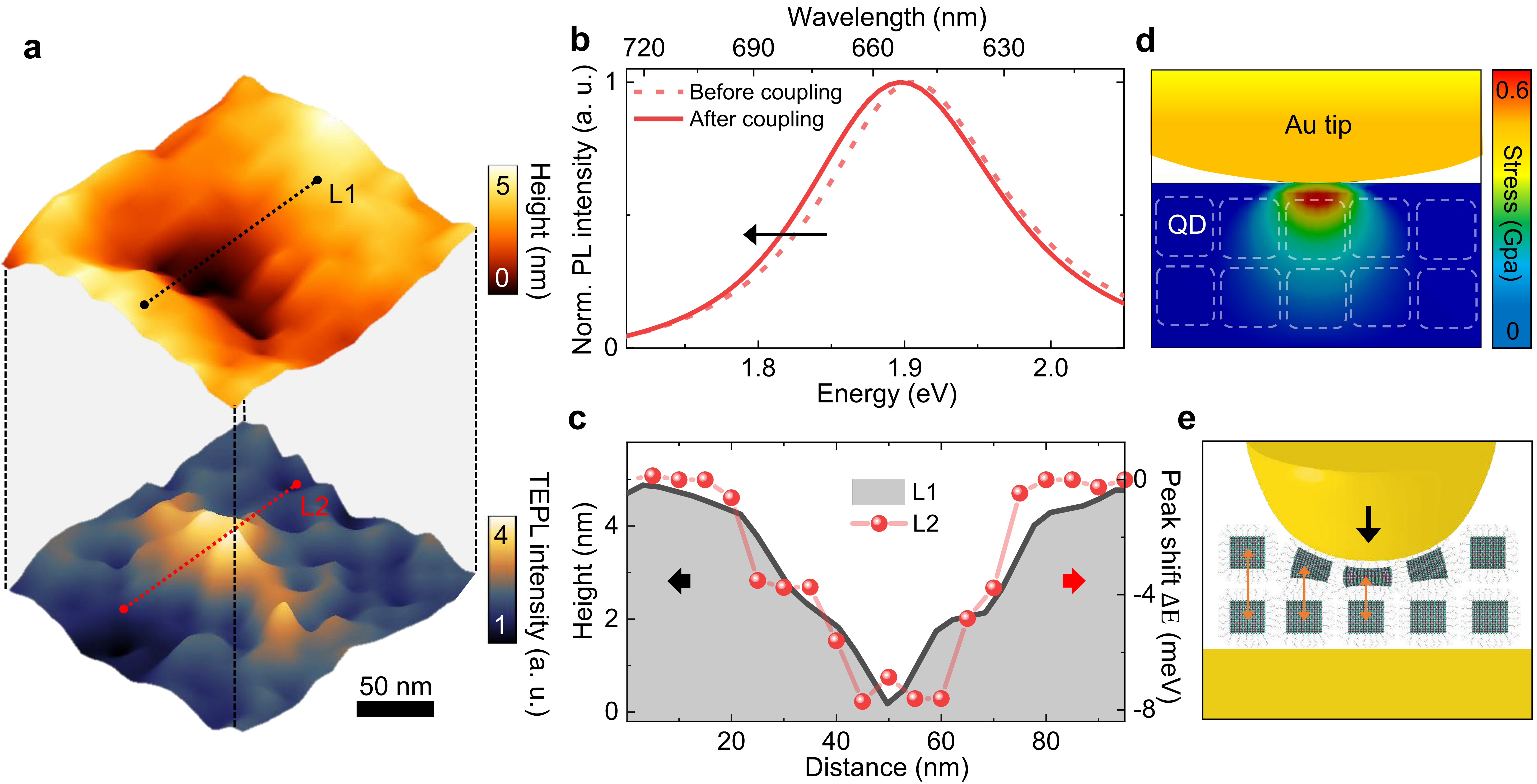}
	\caption{
\textbf{Nanoscale heterogeneity in the pQD ensemble.} 
(a) AFM image (top) of the structurally modified CsPbBr$_{x}$I$_{3-x}$ pQD ensemble induced by tip pressure and corresponding TEPL peak intensity image (bottom).
(b) TEPL spectra of the pQD ensemble obtained before (red dashed line) and after (red line) pQD coupling.
(c) Topographic line profile (black) derived from line L1 in the AFM image (a, top) and peak energy shift (red) corresponding to line L2 in TEPL image (a, bottom).
(d) Numerical simulation of the stress distribution when the Au tip presses the CsPbBr$_{x}$I$_{3-x}$ pQD ensemble.
(e) Illustration of the facilitated pQD coupling region attributed to the decreased inter-pQDs distance as the Au tip presses the ensemble.
}
	\label{fig:fig4}
\end{figure*}
%%%%%%%%%%%%%%%%%%%%%%%%%%%%%%%%%%%%%%%%%%%%%

\noindent
\textbf{Deterministic generation of nanoscale heterogeneity in pQD ensemble}

\begin{figure*}
	\includegraphics[width = 14 cm]{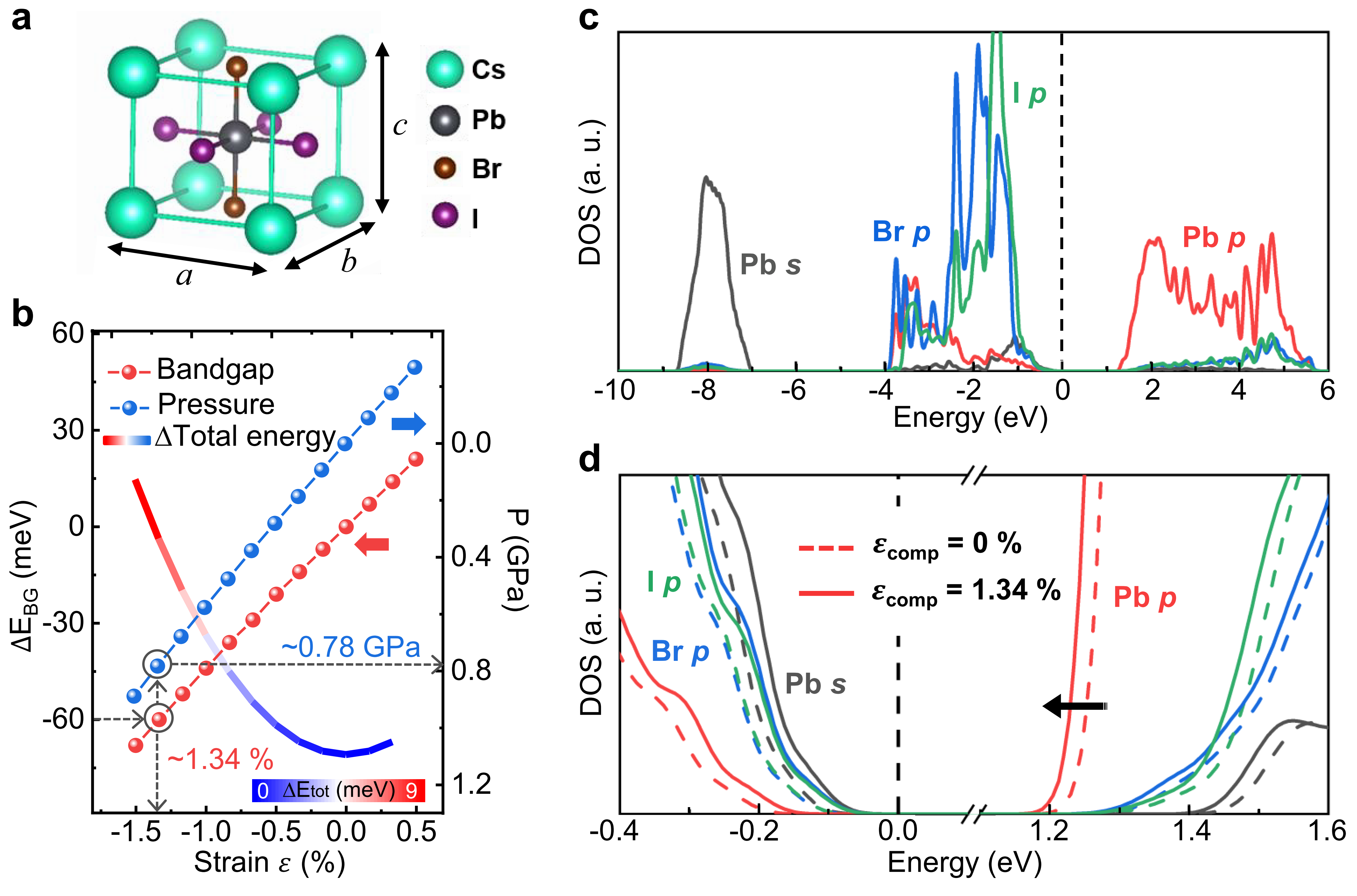}
	\caption{
		\textbf{Electronic and optical properties of mechanically strained CsPbBrI$_{2}$ perovskite.} 
		(a) Atomic structure of CsPbBrI$_{2}$ perovskite. (b) Bandgap shift and applied pressure depending on the c-axis strain. The free energy graph is independently plotted in the background. (c) Projected density of states (PDOS) for the Pb \textit{s} orbital and Pb, Br, and I \textit{p} orbitals of CsPbBrI$_{2}$ equilibrium structure. The dotted line indicates the Fermi energy. (d) PDOS for the \textit{s} and \textit{p} states of Pb and \textit{p} states of Br and I near the Fermi energy under 0 $\%$ (dashed line) and 1.34 $\%$ (solid line) strains. The two energy scales are aligned with respect to their core level.
			}
	\label{fig:fig5}
\end{figure*}

\noindent 
As an extension of the tip-induced control study, we press the CsPbBr$_{x}$I$_{3-x}$ pQD ensemble with the Au tip to investigate the interactive effect between pQDs under applied local pressure (see Fig. S2 for more details). 
When we compress and decompress the CsPbBr$_{x}$I$_{3-x}$ pQD ensemble, it shows an apparent topographic change in the pressed region, as shown in Fig.~\ref{fig:fig4}a (top).
Simultaneously, we obtain a corresponding TEPL image (bottom of Fig.~\ref{fig:fig4}a) to investigate the optical characteristics of the modified pQD ensemble. 
The TEPL image shows the increased emission intensity at the tip-indented region, and this intensity change is correlated with the topographic change, as shown in Fig.~\ref{fig:fig4}a.
This correlation is understood by the enhanced PLQY as the cavity mode volume, i.e., distance between the Au tip and Au surface, is decreased \cite{park2016}. 
When we compare the TEPL spectra before and after the Au tip presses the CsPbBr$_{x}$I$_{3-x}$ pQD ensemble, they exhibit a distinct spectral redshift with a linewidth broadening, as shown in Fig.~\ref{fig:fig4}b.
To check the correlation between the pressure and spectral redshift, we derive the topographic line profile and peak energy shift profile from the corresponding locations of lines L1 and L2 in Fig.~\ref{fig:fig4}a.
As shown in Fig.~\ref{fig:fig4}c, the size of the tip-indented hole is comparable to the apex size of the Au tip (see Fig. S3 for more details). 
The magnitude of the redshifted energy changes proportionally with respect to the distortion of the ensemble surface, which is confirmed by the simulated stress distribution.
As shown in Fig.~\ref{fig:fig4}d, the stress generated from the Au tip is distributed across the broad region of the pQD ensemble.
Specifically, the maximum-stress region is located on the right underneath of the Au tip, and the induced stress is gradually decreased as the distance increases from the maximum region.
This stress distribution shows that a significant decrease in the inter-pQD distance occurs at the center, and a relatively smaller distance decrease is expected at the near regions, as illustrated in Fig.~\ref{fig:fig4}e.
These observed nano-optomechanical characteristics are in good agreement with previous studies on pQD coupling based on the same mechanism that high pressure causes a significant reduction in the inter-pQD distance \cite{cui2019, li2019, williams2009}.
The reduced inter-pQD distance then leads to pQD coupling with characteristics of reduced energy for exciton transition and spectral broadening by forming multiple states in the coupled pQDs \cite{cui2019, li2019}. 
Therefore, our tip-induced local pressure of $>$0.8 GPa (as shown in Fig.~\ref{fig:fig5}b) can effectively reduce the inter-pQD distance and possibly facilitate pQD coupling in the nanoscale region.
\\

%%%%%%%%%%%%%%%%%%%%%%%%%%%%%%%%%%%%%%%%%%%%%
%%%%%%%%%%%%%%%%%%%%%%%%%%%%%%%%%%%%%%%%%%%%

\noindent
\textbf{DFT calculations for strained CsPbBrI$_{2}$ perovskite}

To understand the electronic and optical properties of mechanically strained pQDs, we perform first-principles DFT calculations on uniaxially-strained CsPbBrI$_{2}$ perovskite (Fig.~\ref{fig:fig5}a), which is compositionally very close to the pQDs used in our experiments (see Fig. S4 in Supplementary Information for more details). 
We obtain the equilibrium lattice constants \textit{a} and \textit{b} at a specific \textit{c} by finding equilibrium volumes using the Murnaghan equation of state \cite{murnaghan1944} from DFT calculations, which give rise to a Poisson’s ratio of 0.143. 
The applied stress is obtained by differentiating the total energy curve as a function of unit cell volume in Fig.~\ref{fig:fig5}b. 
We find that the bandgap decreases almost linearly as the compressive strain increases along the \textit{c}-axis. Fig.~\ref{fig:fig5}b shows the  shift and applied stress depending on the \textit{c}-axis strain. The  reduction of $\sim$60 meV corresponds to 1.34 $\%$ compressive strain. 
Thus, we derive a compressive stress of 0.78 GPa applied to the CsPbBrI$_{2}$ structure.  
Fig.~\ref{fig:fig5}c shows the overall density of states (DOS) of CsPbBrI$_{2}$ perovskite. 
The valence band maximum (VBM) is derived from the \textit{p} states of Br and I with contribution from the Pb \textit{s} states. 
Whereas, the conduction band minimum (CBM) predominantly consists of the \textit{p} orbital of Pb. 
Under the compressive strain of 1.34 $\%$, the \textit{s} orbital of Pb and \textit{p} orbital of the halide are up-shifted in the valance band edge, while the \textit{p} orbital of Pb in the conduction band edge is down-shifted, resulting in the  decrease (Fig.~\ref{fig:fig5}d). 
Specifically, as the CsPbBrI$_{2}$ crystal is compressed by 1.34 $\%$ along the \textit{c}-axis, the in-plane Pb$\--$I bond length increases by 0.006 \AA, while the Pb$\--$Br bond length decreases by 0.040 \AA. 
Because the VBM and CBM are the antibonding states of the Pb \textit{s} and halide \textit{p} orbitals and the Pb \textit{p} and halide \textit{p} orbitals, respectively \cite{filippetti2014, brandt2015, ravi2016, brandt2015, umebayashi2003}, the CBM moves downwards alongside the increased Pb$\--$I bond lengths, and the VBM moves upwards alongside the decreased Pb$\--$Br bond lengths (see Fig. S5 in Supplementary Information for more details).
\\

\noindent
{\bf Conclusions}

\noindent
In summary, we have demonstrated the tip-induced dynamic control of strain, bandgap, and quantum yield of single CsPbBr$_{x}$I$_{3-x}$ pQDs by using a controllable plasmonic nano-cavity combined with TEPL spectroscopy.
This tip-induced nano-engineering approach with local pressure of GPa order not only provides an in-depth understanding of the optical properties of metal halide perovskite pQDs and their coupling nature but also offers a practical way to tune the mechanical and electronic properties at the single-pQD level for potential applications in ultracompact nano-optoelectronic devices.
Specifically, we envision that this dynamical single-dot manipulation will enable the tunable nano-LEDs \cite{liu2014}, realizing ultra-high definition display with high efficiency.
In addition, extremely high tip induced local pressure will allow pressure-induced recrystallization and phase transition at the few-nanometer-length scale to further improve the grain quality and optical properties of 2D perovskite ensembles as a post-fabrication process.
Furthermore, the extrinsically modified physical properties of pQDs by the plasmonic cavity, e.g., reduced non-radiative recombination, modulation of carrier dynamics and PL lifetime, and enhanced radiative decay rate, can significantly improve the efficiency of perovskite photovoltaics \cite{kim2018, wang2017}.
\\

\noindent
{\bf Methods}

\noindent
\textbf{Synthesis of CsPbBr$_x$I$_{3-x}$ pQDs.}
Cesium carbonate (Cs$_2$CO$_3$, 99.9 $\%$, Sigma-Aldrich), lead iodide (PbI$_2$, 99.99 $\%$, TCI), oleic acid (OA, technical grade, Sigma-Aldrich), oleylamine (OLA, technical grade, Sigma-Aldrich), zinc bromide (ZnBr$_2$, 99.999 $\%$, Sigma-Aldrich), and 1-octadecene (ODE, technical grade, Sigma-Aldrich) were used. Cesium oleate precursor solution was prepared based on previously reported procedures \cite{protesescu2015, woo2017} with modifications. 0.163 g of Cs$_2$CO$_3$, 0.5 mL of OA, and 8 mL of ODE were placed in a round-bottomed 3-neck flask and heated to 110 $^{\circ}\mathrm{C}$ under vacuum. After degassing for 2 h, the solution was further heated to 150 $^{\circ}\mathrm{C}$ under an inert atmosphere. Also, 0.347 g of PbI$_2$, 0.085 g of ZnBr$_2$, 2.0 mL of OA, 2.0 mL of OLA, and 20 mL of ODE were placed in a round-bottomed 3-neck flask and degassed at 110 $^{\circ}\mathrm{C}$ for 2 h. Then, the solution was heated to 180 $^{\circ}\mathrm{C}$ under an inert environment. Finally, 1.6 mL of preheated cesium oleate precursor solution was quickly injected, and the reaction solution was rapidly quenched in an ice bath. The pQDs were collected from the crude reaction solution after multiple purification steps. Specifically, the crude reaction solution was centrifuged at 13000 rpm for 30 min, and the precipitates were redispersed in 6 mL anhydrous toluene. The pQD toluene solution was centrifuged (6000 rpm for 5 min) with an additional 4 mL of methyl acetate, and the precipitates were redispersed in 8 mL of anhydrous hexane. The QD hexane solution was again centrifuged at 6000 rpm for 5 min. Finally, the supernatant was collected and filtered using a Teflon syringe filter (0.45 micron). Note that all procedures were conducted in an inert atmosphere.
\\

\noindent
\textbf{TEPL spectroscopy setup.}
The prepared pQDs spin-coated on the template stripped gold substrate were loaded on a piezo-electric transducer (PZT, P-611.3X, Physik Instrumente) for XY scanning and application of compressive strain to single pQDs with $<$0.2 nm positioning precision. To directly press the single pQDs, an Au tip (apex radius of $\sim$10 nm) was used. The Au tip, which was fabricated with a refined electrochemical etching protocol, was attached to a quartz tuning fork (resonance frequency of 32.768 kHz) to regulate the distance between the tip and sample based on shear-force AFM operated by a digital AFM controller (R9$+$, RHK Technology). For the TEPL experiments, a conventional optical spectroscopy setup was combined with home-built shear-force AFM. For a high-quality wavefront of the excitation beam, a He$\--$Ne laser (632.8 nm, $<$0.5 mW) was coupled and passed through a single-mode fiber (core diameter of $\sim$3.5 $\mu$m) and collimated again using an aspheric lens. The collimated beam was then passed through a half-wave plate to make the excitation polarization parallel with respect to the tip axis. Finally, the beam was focused onto the Au tip using a microscope objective (NA $=$ 0.8, LMPLFLN100X, Olympus) with a side illumination geometry. To ensure highly efficient laser coupling to the Au tip, the tip position was controlled with $\sim$30 nm precision by Picomotor actuators (9062-XYZ-PPP-M, Newport). TEPL responses were collected using the same microscope objective (backscattering geometry) and passed through an edge filter (LP02-633RE-25, Semrock) to cutoff the fundamental laser line. TEPL signals were then dispersed onto a spectrometer (f $=$ 328 mm, Kymera 328i, Andor) and imaged with a thermoelectrically cooled charge-coupled device (CCD, iDus 420, Andor) to obtain the TEPL spectra. Before the experiment, the spectrometer was calibrated with an Argon Mercury lamp. A 150g$/$mm grating blazed to 800 nm (spectral resolution of 0.62 nm) was used for PL measurements.
\\

\noindent
\textbf{DFT calculations.}
For the first-principles DFT calculations, we employed the projector-augmented wave potentials \cite{blochl1994} with ionic core potential and the Perdew-Burke-Ernzernhof exchange-correlation functional \cite{perdew1996}, as implemented in the Vienna \textit{Ab-initio} Simulation Package (VASP) \cite{kresse1999}. We used a kinetic energy cutoff of 400 eV and an ($8\times8\times8$) K-points grid for sampling the $\Gamma$-point. Gaussian smearing of 0.05 eV was employed. All atomic forces were relaxed to less than 0.01 eV$/$\AA.
\\

\noindent
\textbf{FDTD simulations of Purcell factor and optical field distribution.}
We used a commercial FDTD simulation software (Lumerical Solutions, Inc.) to characterize the Purcell enhancement factor and optical field distribution at the Au tip$\--$Au substrate nano-gap. The Au tip with a 5 nm apex radius was placed in close proximity (10 nm gap) to the Au substrate. As a fundamental excitation source, a linearly polarized monochromatic 633 nm light was projected with 45$^{\circ}$. For estimation of the Purcell enhancement factor, a fluorescent dipole with emission wavelength 650 nm was positioned at the center of the nano-gap.
\\

\noindent
\textbf{Numerical analysis of the tip-induced stress distribution.}
A three-dimensional mechanical simulation was performed to analyze the distribution of the pQD ensemble and the pressure applied to the contact region between the Au tip and pQDs. The pressure and force were calculated through the ANSYS Mechanical Enterprise module. The lowest position of the Au tip was set as the position of the surface of the pQDs to model the experimental conditions. The bottom plane of the pQDs was considered to be fixed at the substrate surface because the pQDs were rigidly covered by a thin Al$_{2}$O$_{3}$ capping layer supressing their lateral movement. Then, the mechanical properties, including stress and pressure, were simulated under gradual increases in the force applied to the Au tip. The applied force was defined on the top surface of the Au tip in the perpendicular direction with respect to the pQD surface. The mechanical properties of Young’s modulus, Poisson ratio, and density of the pQDs were obtained from the literature \cite{rakita2015}.
\\

\noindent
\textbf{Data availability.}
The data that support the plots within this paper and other findings of this study are available from the corresponding author upon reasonable request.

\bibliography{perov} 

\vskip 1cm
\noindent
{\bf Acknowledgements}

\noindent
This work was supported by a National Research Foundation of Korea (NRF) grant funded by the Korea government (MEST) (2019K2A9A1A06099937 and 2020R1C1C1011301). S.J. acknowledges a Creative Materials Discovery Program through the National Research Foundation (NRF) of Korea funded by the Ministry of Science and ICT (NRF-2019M3D1A1078299, 2019R1A2B5B03070407). Y.-H.K acknowledges 2019M3D1A1078302.

\noindent
{\bf Author contributions}

\noindent
H. Lee and K.-D Park conceived the experiments.
H. Lee performed the TEPL spectroscopy and control experiments.
J.Y. Woo, J. Park, and S. Jeong designed and prepared the samples.
I. Jo, Y. Lee, and Y.-H Kim designed and performed the simulations.
Y. Koo, D.Y. Park, and H. Kim performed pre-characterizations with the PL and AFM measurements.
J. Choi designed and developed the hyperspectral imaing processes.
H. Lee, J.Y. Woo, S. Jeong, and K.-D. Park analyzed the data, and all authors discussed the results.
H. Lee and K.-D. Park wrote the manuscript with contributions from all authors.
K.-D. Park supervised the project.

\noindent
{\bf Additional information}

\noindent
Supplementary information is available for this paper.

\noindent
{\bf Competing financial interests}

\noindent
The authors declare no competing financial interests.

\end{document}